 \documentclass[10pt,pre,twocolumn,amsmath,amssymb,floatfix]{revtex4} 
 \usepackage{bm}
 \usepackage{graphicx, hyperref}
 \usepackage{latexsym}

 \newcommand{\boldxi}{\mbox{\boldmath$\xi$\unboldmath}}
 \newcommand{\boldF}{\mbox{\boldmath$\cal F$\unboldmath}}

 \newcommand{\iotabar}{\mbox{$\,\iota\!\!$-}}
 \newcommand{\Fig}[1]{Fig.~\ref{fig:#1}}
 \newcommand{\Tab}[1]{Table.~\ref{tab:#1}}
 \newcommand{\Eqn}[1]{Eq.\,(\ref{eq:#1})}
 \newcommand{\Sec}[1]{Sec.\,\ref{sec:#1}}

 \newcommand{\be}{\begin{eqnarray}}
 \newcommand{\ee}{\end{eqnarray}}

 \newcommand{\Poincare}{Poincar{\'e} }

 \newcommand{\insertfigure}[3]{
  \begin{figure}[h]
  \begin{center}
   \includegraphics{#1}
  \caption{#2\label{fig:#3}}
  \end{center}
  \end{figure}
  }

 \newcommand{\insertdblfigure}[3]{
 \begin{figure*}[t]
 \begin{center}
 \includegraphics{#1}
 \caption{#2\label{fig:#3}}
 \end{center}
 \end{figure*}
 }

\begin{document}

 \title{Nonaxisymmetric, multi-region relaxed magnetohydrodynamic equilibrium solutions}

 \author{S.R. Hudson}
 \email{shudson@pppl.gov}

 \affiliation{Princeton Plasma Physics Laboratory, PO Box 451, Princeton NJ 08543, USA}

 \author{R.L. Dewar}
 \email{robert.dewar@anu.edu.au}

 \author{M.J. Hole}
 \email{matthew.hole@anu.edu.au}

 \author{M. McGann}
 \email{mathew.mcgann@anu.edu.au}

 \affiliation{Plasma Research Laboratory, Research School of Physics \& Engineering, The Australian National University, Canberra, ACT 0200, Australia}

 \date{\today}

 \begin{abstract}
 We describe a magnetohydrodynamic (MHD) constrained energy functional for equilibrium calculations that combines the topological constraints of ideal MHD with elements of Taylor relaxation.
 Extremizing states allow for partially chaotic magnetic fields and non-trivial pressure profiles supported by a discrete set of ideal interfaces with irrational rotational transforms.
 Numerical solutions are computed using the Stepped Pressure Equilibrium Code, and benchmarks and convergence calculations are presented.
 \end{abstract}

 \maketitle

 \section{introduction}

 The construction of magnetohydrodynamic (MHD) equilibria in three-dimensional (3D) configurations is of fundamental importance for understanding magnetically confined plasmas.
 To illustrate both the importance and subtlety of this problem: it is widely accepted that quiescent plasma confinement depends on constructing equilibria that are stable to small perturbations,
 which necessarily presupposes the existence of an equilibrium;
 however, as pointed out by Grad \cite{Grad_67}, ``a more primitive reason than instability for the lack of confinement is the absence of an appropriate equilibrium state''.

 Given that all experimental confinement devices lack a continuous symmetry to some extent, either slightly so because of discrete coil effects, error fields, or intentionally applied resonant magnetic perturbations \cite{Evans_Moyer_Thomas_etal_04}, or greatly so because of intrinsic 3D shaping such as in the stellarator class \cite{Spitzer_58,Boozer_98} of confinement devices, the computation of 3D equilibrium solutions with arbitrarily chaotic fields is of foremost importance.

 The theory and numerical construction of 3D equilibria is greatly complicated by the fact that toroidal magnetic fields without a continuous symmetry are generally a fractal mix of islands, chaotic field lines, and magnetic flux surfaces.
 Any deformation of the plasma boundary, or error field, that resonates with rational field lines will generally (in the absence of ideal surface currents at the rational surfaces) result in the formation of magnetic islands.
 Where these islands overlap, regions of connected chaos, so-called stochastic volumes, will form \cite{Chirikov_79}.
 According to Greene \cite{Greene_79}, ``there is a stochastic region in the immediate vicinity of every chain of periodic orbits''.
 In contrast, the Kolmogorov--Arnold--Moser theorem indicates that flux surfaces with ``sufficiently irrational'' rotational transform can survive small perturbation \cite{Moser_73,Arnold_78}.
 The rotational transform, $\iotabar$, is considered sufficiently irrational if it satisfies a Diophantine condition, e.g. \mbox{$|\iotabar-n/m| > r m^{-k}$} for all integers \mbox{$n$} and \mbox{$m$}, and where \mbox{$r>0$} and \mbox{$k\ge 2$}.
 
 \insertfigure{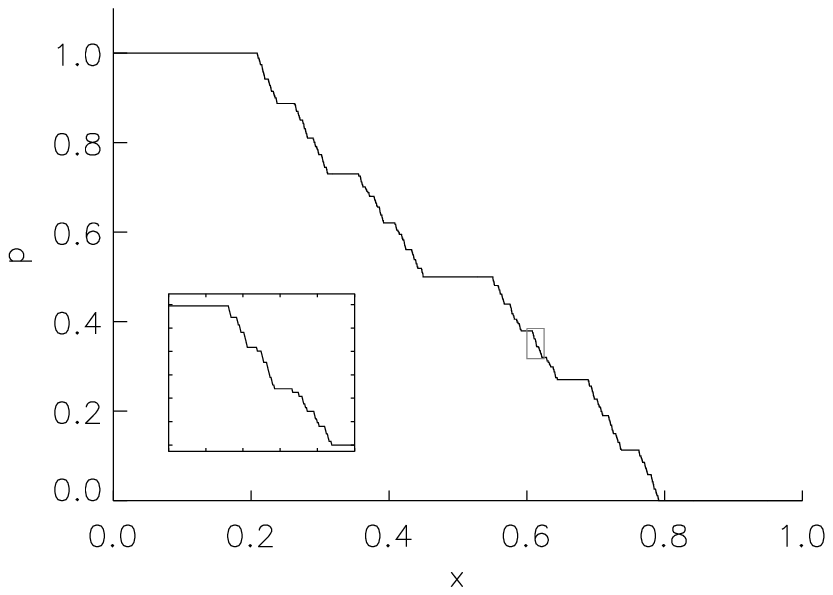}{Diophantine pressure profile with $r=0.2$ and $k=2$, normalized so that $p(0)=1$ and $p(1)=0$. The large rectangular region is a magnification of the small rectangular region.}{DiophantinePressure}

 The fractal phase-space structure of non-symmetric, and therefore generally non-integrable (i.e. such that a continuously nested family of flux surfaces does not exist), magnetic fields has important consequences for the construction of scalar-pressure ideal equilibria, i.e. solutions to \mbox{$\nabla p = {\bf j}\times{\bf B}$}.
 Ideal force balance immediately requires that \mbox{${\bf B}\cdot \nabla p=0$}, so that the pressure is constant along a field line, which in turn implies that the pressure must be constant in the stochastic volumes.
 Pressure gradients can be supported on the flux surfaces that survive perturbation, but any non-trivial continuous pressure profile that is consistent with \mbox{$\nabla p = {\bf j}\times{\bf B}$} {\em and} the fractal structure of a non-integrable field must have a gradient that is everywhere discontinuous or zero \cite{Hudson_Nakajima_10} and is something akin to a devil's staircase.
 For example, consider the pressure profile, $p(x)$, defined by $p'(x)=1$ if \mbox{$|x-n/m| > r m^{-k}$} for all rationals $n/m$, and $p'(x)=0$ otherwise.
 The function $p'(x)$ does not have a well defined Riemann integral, but an approximation to $p(x)$ is shown in \Fig{DiophantinePressure}.
 
 A continuous profile may be constructed, but the gradient is infinitely discontinuous.
 This immediately causes problems for the construction of scalar-pressure equilibria, as it is the pressure gradient, $\nabla p$, rather than the pressure itself, that appears in \mbox{$\nabla p = {\bf j}\times{\bf B}$}.
 Grad went on to conclude that non-trivial equilibria must have ``very pathological pressure'' \cite{Grad_67}.
 The pathological structure of scalar-pressure, ideal MHD equilibria with chaotic fields causes problems for existing numerical algorithms {\cite{Hudson_Nakajima_10}.

 The most elegant approaches to numerical construction of equilibrium solutions employ energy principles, and this approach began to be developed as a practical method for numerical calculation of 3D ideal-MHD equilibria quite early in the history of computational plasma physics \cite{Betancourt_Garabedian_76,Hirshman_Whitson_83,Bhattacharjee_Wiley_Dewar_84,Lao_etal_85}. Today the VMEC code  \cite{Hirshman_Rij_Merkel_86}, which has been continuously developed since that time, has become the most widely used 3D equilibrium code.
 
Such codes seek solutions that minimize the plasma energy, \mbox{$U = \int [p_l/(\gamma-1)+B^2/2\mu_0]dv$}, but if this functional is minimized allowing for arbitrary variations without constraints then the minimizing state is trivial \cite{Kruskal_Kulsrud_58}. Instead, in ideal-MHD codes, the plasma energy is minimized under the continuous \emph{infinity} of constraints implied by the ideal-MHD equations at each point in the plasma, e.g. \mbox{$\delta {\bf B}=\nabla \times(\boldxi \times {\bf B})$}, representing frozen-in flux in each plasma element.
However, such variations do not allow the topology of magnetic surfaces to change, so that reconnection and relaxation phenomena are precluded.
If the topology assumed in the initial guess for the equilibrium is that of a nested family of flux surfaces foliating the plasma domain, then it is constrained to remain so during the energy minimization, so that the generic state for a 3D equilbrium, one with magnetic islands and chaos, cannot be obtained in these ideal codes, limiting the attainable precision because of the formation of unresolved current singularities.

Thus we need to reduce the number of constraints if we are to have a well-posed energy principle for 3D equilibria, the most extreme reduction being provided by \emph{Taylor relaxation} \cite{Taylor_74}.
 Allowing for arbitrarily small resistivity and arbitrary variations, so that the topological constraints on the plasma evolution are removed, Taylor argued that a weakly resistive plasma will relax to minimize the plasma energy subject to the constraint of conserved helicity, $H = \int {\bf A} \cdot {\bf B} dv$. 
 The resulting state is a force-free field, and so a globally relaxed Taylor state cannot support a pressure gradient ---
 the magnetic field topology is not constrained but the pressure is constant everywhere,  suggesting that a few more constraints are required to model a confined plasma satisfactorily.

Augmenting the helicity constraint with one or more \emph{smooth} moments of ${\bf A} \cdot {\bf B}$ leads to a practical approach to constructing axisymmetric equilibria \cite{Bhattacharjee_Wiley_83,BGAS_86} with non-trivial pressure profile and low free energy to model a toroidally confined plasma with good stability properties.

However this approach requires good magnetic surfaces everywhere in the plasma and is thus limited to integrable magnetic fields. For the 3D equilibrium problem we are forced to consider \emph{non-smooth} constraints \cite{DHMMH_08}, giving an energy principle that allows for equilibria with partially chaotic fields and non-trivial, but discontinuous, pressure profile.
 This model, which we call multi-region, relaxed magnetohydrodyamics, or MRXMHD, combines ideal topological constraints at a discrete set of selected irrational surfaces with Taylor relaxation in between. This is a partial relaxation model so that non-zero pressure gradients are supported at the selected irrational surfaces but topological reconnection is possible at the intervening rational surfaces.
 This model is described in \Sec{MRXMHD}, and in \Sec{examples} we present some illustrations of MRXMHD equilibrium solutions, as computed by the newly implemented Stepped Pressure Equilibrium Code, SPEC.

  \section{multi-region, relaxed MHD}\label{sec:MRXMHD}

 The plasma is modeled \cite{DHMMH_08} as a collection of nested annular regions, \mbox{${\cal V}_l$} for \mbox{$l=1,..,N$}, which are separated by a discrete set of toroidal surfaces, \mbox{${\cal I}_l$}, so that \mbox{${\cal V}_{l}$} is bounded by \mbox{${\cal I}_{l-1}$} and \mbox{${\cal I}_{l}$}.
 In each ${\cal V}_l$, the magnetic field relaxes to minimize the plasma energy, subject to the constraint of conserved helicity and mass/entropy.
 On each \mbox{${\cal I}_l$}, we apply the constraints of ideal MHD.
 Equilibrium states are extrema of the constrained energy functional,
 \be F = \sum_l \left( U_l - \mu_l H_l / 2 - \nu_l M_l /2\right), \label{eq:constrainedenergyfunctional}
 \ee
 where the plasma energy, helicity and mass/entropy in each annulus are given respectively by 
 \be U_l & = & \int_{{\cal V}_l} \left(\frac{p_l}{\gamma-1}+\frac{B_l^2}{2\mu_0}\right)dv, \label{eq:energy}\\
     H_l & = & \int_{{\cal V}_l} {\bf A}_l\cdot{\bf B}_l\,dv, \label{eq:helicity}\\
     M_l & = & \int_{{\cal V}_l} p_l^{1/\gamma}dv.\label{eq:mass}
\ee 
 The magnetic field is described by a vector potential, \mbox{${\bf B}_l=\nabla\times{\bf A}_l$}, the \mbox{$\mu_l$} and \mbox{$\nu_l$} are Lagrange multipliers, and \mbox{$\gamma$} is the ratio of specific heats.
 Hereafter, the permeability of free space factor, \mbox{$\mu_0$}, will be omitted.

 In extremizing \mbox{$F$} we allow for arbitrary variations in the pressure, \mbox{$\delta p_l$}, and the vector potential, \mbox{$\delta {\bf A}_l$}, in each annulus, and the geometry, \mbox{$\boldxi_l$}, of the toroidal surfaces.
 To enforce the constraint that the magnetic field remain tangential to the toroidal surfaces, on the \mbox{${\cal I}_l$} the variations in the field are related to the variations in geometry by using a gauge such that \mbox{$\delta {\bf A} = \boldxi \times {\bf B}$}.
 We appropriately call these toroidal surfaces {\em ideal interfaces}.

The first order variation in the free-energy functional in each annulus is
\be \delta F_l & = & \int_{{\cal V}_l} \left(\frac{1}{\gamma-1} - \nu_l \frac{p_l^{1/\gamma}}{\gamma p_l}\right) \delta p_l \; dv \nonumber \\
               & + & \int_{{\cal V}_l} \left( \nabla \times {\bf B}_l-\mu_l {\bf B}_l \right) \cdot \delta {\bf A}_l \; dv \label{eq:firstvariation} \\
               & + & \int_{\partial {\cal V}_l} \left( \frac{p_l}{\gamma-1} - \nu_l p_l^{1/\gamma} - \frac{B_l^2}{2} \right) ({\bf n}\cdot \boldxi ) \; ds
\nonumber 
\ee
 The equilibrium is thus comprised of a family of nested Beltrami fields, where \mbox{$\nabla \times {\bf B}_l = \mu_l {\bf B}_l$} in each annulus, the pressure is constant in each annulus, and the total pressure is continuous across the ideal interfaces, \mbox{$[[p+B^2/2]]=0$}.

 Pressure is supported by the ideal interfaces, across which a pressure discontinuity is allowed provided there is a compensating discontinuity in the tangential field.
 The equilibrium solutions are topologically-constrained, partially relaxed, stepped-pressure states.
 Topological constraints on magnetic reconnection have been observed in a similar context \cite{YHW_10}.
 A strong motivation for adopting this model is that Bruno \& Laurence \cite{Bruno_Laurence_96} have proved that, under certain conditions, such equilibria exist.
 This places the MRXMHD model on a strong mathematical foundation.

 In addition to a prescribed boundary, we define the equilibrium by prescribing pressure and rotational-transform profiles, given as functions of the toroidal flux as described below.
 
 The pressure is constant in each annulus but the rotational transform, well-defined on flux surfaces and defined by suitable interpolation across chaotic regions, changes across the annuli. 
 Given that the topology of the field is arbitrary within the annular regions, and that only the interfaces are guaranteed to be flux surfaces, the rotational transform is prescribed {\em only} on the ideal interfaces.
 To avoid discontinuous rotational-transform profiles we require the rotational transforms to be the same on each side of an interface.
 So, to specify the pressure and rotational-transform profiles, we give the toroidal flux and the pressure, $p_l$, in each annulus, and the rotational transform, $\iotabar_l$, on each interface, for $l=1,N$.
 We hold the toroidal flux in each annulus fixed (though for reversed-field pinches it would be more appropriate to fix the poloidal fluxes).
 Also, as is typically done in equilibrium calculations, we hold the $p_l$ and $\iotabar_l$ constant throughout the calculation.
 However, it should be noted that this is not consistent with the variational principle, so the constraints Eqs.~(\ref{eq:helicity}) and (\ref{eq:mass}) must be adjusted iteratively during the calculation.

 An initial guess for the geometry of the ideal interfaces, ${\cal I}_l$, is given in cylindrical coordinates, \mbox{$(R,\phi,Z)$}, via
 \mbox{$R_l(\theta,\zeta) = \sum_j R_{l,j} \cos(m_j\theta-n_j\zeta)$} and 
 \mbox{$Z_l(\theta,\zeta) = \sum_j Z_{l,j} \sin(m_j\theta-n_j\zeta)$},
 where $\theta$ is at this stage an arbitrary poloidal angle and \mbox{$\zeta=-\phi$}.
 (We have restricted attention to stellarator-symmetric configurations \cite{Dewar_Hudson_97b}.)
 The toroidal and poloidal fluxes, \mbox{$\psi_{t,l}$} and $\psi_{p,l}$, enclosed by each interface are also assumed given.
 A piecewise-cubic interpolation of the interfaces using the radial coordinate \mbox{$s\equiv\sqrt \psi_{t}$} provides a smooth, global coordinate system, \mbox{$(s,\theta,\zeta)$}, 
 with coordinate Jacobian $\sqrt g=(\nabla s \cdot \nabla \theta \times \nabla \zeta)^{-1}$.

 The vector potential in each annulus is written \mbox{${\bf A}=A_\theta \nabla \theta + A_\zeta \nabla \zeta$}, and \mbox{$A_\theta$} and \mbox{$A_\zeta$} are discretized using a mixed Fourier, finite-element representation, e.g. \mbox{$A_\theta = \sum_{j} A_{\theta,j}(s)\cos(m_j\theta-n_j\zeta)$}, and the radial dependence is described by \mbox{$A_{\theta,j}(s)=\sum_{k}A_{\theta,j,k}\varphi_k(s)$}, where the \mbox{$\varphi_k(s)$} are piecewise-quintic basis functions with finite support, defined on a radial sub-grid (an example of which is shown below).
 The $A_{\theta,j,k}$ and $A_{\zeta,j,k}$ are constrained to ensure that the flux constraints are satisfied, and that \mbox{$\sqrt g {\bf B} \cdot \nabla s \equiv \partial_\theta A_\zeta - \partial_\zeta A_\theta=0$} at the interfaces, but is otherwise general.
 Setting the derivatives of \Eqn{constrainedenergyfunctional} with respect to the $A_{\theta,j,k}$ and $A_{\zeta,j,k}$ to zero allows each ${\bf B}_l$ to be efficiently determined as the solution to a sparse system of linear equations.
 Each ${\bf B}_l$ depends only on the geometry of bounding interfaces, ${\cal I}_{l-1}$ and ${\cal I}_l$, the enclosed fluxes, and the Lagrange multiplier, \mbox{$\mu_l$}, which is related to the parallel current.
 This must be adjusted to ensure that the rotational transform on the ideal interfaces matches the prescribed value.
 The computation of the Beltrami fields in multiple regions is trivially distributed across multiple processors.

 The innermost volume contains the coordinate origin, where the coordinate Jacobian goes to zero.
 At the origin, we enforce the condition that the geometry of the interfaces is regular, and the geometry of the innermost interface is obtained by extrapolation.

 The interface geometry must be adjusted in order to satisfy force balance, \mbox{$[[p+B^2/2]]=0$}.
 The first-order variation in the energy functional, $F$, depends only the {\em normal} component of the geometrical variation, \mbox{$\boldxi \cdot {\bf n}$}.
 In order to obtain a unique Fourier representation of the interface geometry, we follow the approach used in VMEC \cite{Hirshman_Breslau_98} and construct an angle that minimizes a measure of the spectral width, \mbox{$\sum_j (m_j^p+n_j^q)(R_j^2+Z_j^2)$}, where \mbox{$p$} and \mbox{$q$} are arbitrary integers controlling the degree of spectral condensation (in the following section we choose $p=4$ and $q=4$), and so obtain an optimally accurate representation of the interface geometry with a finite set of Fourier harmonics.
 Allowing for tangential variations of the form, \mbox{$\delta R = \partial_\theta R \, \delta u$} and \mbox{$\delta Z = \partial_\theta Z \, \delta u$}, the condition that $\theta$ minimizes the spectral width of each interface is \mbox{$I_l \equiv \partial_\theta R_l \, X + \partial_\theta Z_l \, Y=0$}, where \mbox{$X \equiv \sum_j (m_j^p+n_j^q) R_{l,j} \cos(m_j\theta-n_j\zeta)$} and \mbox{$Y \equiv \sum_j (m_j^p+n_j^q) Z_{l,j} \sin(m_j\theta-n_j\zeta)$}.

 The interface geometries are described by the $R_{l,j}$ and the $Z_{l,j}$, which we collect together as a vector, ${\bf x}$.
 We construct a vector of constraints, $\boldF$, as a collection of the Fourier harmonics of the force imbalance, $[[p+B^2/2]]_{l,j}$, and the spectral constraints, $I_{l,j}$, at each interface.
 The task of constructing equilibrium solutions is thus reduced to the standard mathematical problem of finding a zero of a multi-dimensional function, $\boldF({\bf x})=0$, which is solved using a mixed Newton, convex-gradient method provided by the NAG library.
 Further details of the algorithm, including convergence studies and benchmark calculations, will be presented elsewhere.

 An analysis of the force-balance condition, \mbox{$[[p+B^2/2]]=0$}, shows that, generally, in order for an interface to support pressure, it must have irrational rotational transforms on each side \cite{MHDvN_10}. As previously mentioned, we take these to be equal.
 ``Noble'' irrationals play an important role in chaos as, typically, flux surfaces with noble rotational transform are locally the most robust \cite{Greene_79,Mackay_Stark_92}. 
 We thus constrain the ideal interfaces that support pressure to have noble rotational transform, given as a Fibonacci-ratio limit: $p_{n+1}/q_{n+1} \equiv (p_{n-1}+p_{n})/(q_{n-1}+q_{n})$ as $n \to \infty$, beginning from any pair of rationals, \mbox{$p_1/q_1$} and \mbox{$p_2/q_2$} which satisfy \mbox{$|p_1 q_2 - p_2 q_1|=1$}.
 The limiting ratio is \mbox{$\iotabar=(p_1+\gamma p_2)/(q_1+\gamma q_2)$}, where \mbox{$\gamma\equiv(1+\sqrt 5)/2$} is the golden mean.
As mentioned earlier, adjustments must be made iteratively during the equilibration calculation.
 Specifically, we adjust the helicity multiplier, $\mu_l$, which is related to the parallel current, and the poloidal flux in each annulus to satisfy the interface rotational-transform constraints.

 \section{non-axisymmetric examples}\label{sec:examples}

 In the following we present numerical examples of stepped-pressure equilibria with 3D boundaries.
 The following examples are constructed by applying a non-axisymmetric deformation to an otherwise axisymmetric configuration.
 The axisymmetric configuration is defined by a fixed boundary with a major radius of $1\,$m and a circular cross section with minor radius $30\,$cm.
 For simplicity of illustration, and computational expediency, we restrict attention to equilibria with only four annular regions, bounded by four interfaces.
 The pressure profile, shown in \Fig{Pressure}, is a piecewise-flat approximation to $p(\psi) = p_0 ( 1-2\psi+\psi^2 )$, where $p_0$ is a scaling factor.
 The interface rotational-transform profile is a discrete approximation to $\iotabar(\psi)=0.8839642543 - 0.7799929021 \psi$, as described in \Tab{interfacetransforms} and shown as the small squares in \Fig{Transform}.

  \begin{table}[h]\caption{Ideal-interface rotational transform} \begin{tabular}{|l|cc|c|}\hline\label{tab:interfacetransforms}
      &                                             &     $\iotabar$       &  $\psi_t$  \\ \hline
$ 1)$ & $ (  6 + \gamma  7 ) / (  7 + \gamma  8 ) $ & $ = 0.8687325\dots $ &  0.0195280 \\
$ 2)$ & $ (  2 + \gamma  3 ) / (  3 + \gamma  4 ) $ & $ = 0.7236068\dots $ &  0.2055884 \\
$ 3)$ & $ (  1 + \gamma  1 ) / (  2 + \gamma  3 ) $ & $ = 0.3819660\dots $ &  0.6435933 \\
$ 4)$ & $ (  1 + \gamma  1 ) / (  9 + \gamma 10 ) $ & $ = 0.1039714\dots $ &  1.0000000 \\
  \hline\end{tabular}\end{table}

 \insertfigure{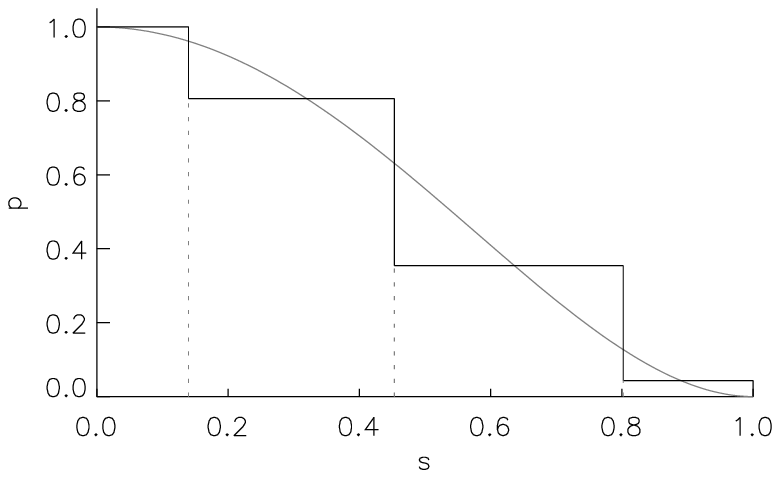}{Smooth pressure profile, $p(\psi)=1-2\psi+\psi^2$, supplied to VMEC, and the stepped pressure profile used in SPEC.}{Pressure}
 \insertfigure{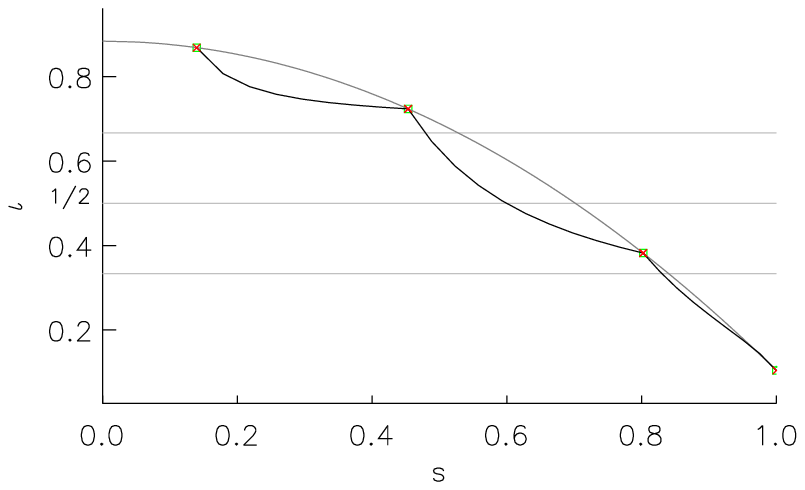}{rotational-transform profile (black line) for the axisymmetric ``high-pressure'' configuration shown in \Fig{VMECComparison}. The small squares indicate where the rotational transform is constrained. The smooth profile (grey line) is supplied to VMEC.}{Transform}

 Below we examine two non-axisymmetric configurations.
 The first is a zero-beta configuration (i.e. $p_0=0$) with an $m=0$, $n=1$ deformation of the boundary.
 This results in an equilibrium with a helical magnetic axis.
 In the second ``high-pressure'' equilibrium, $p_0$ is increased in order to induce a $3\,$cm Shafranov shift of the magnetic axis of the axisymmetric configuration.

 At this point we stress that, despite the fact that only $4$ interfaces were used and the piecewise-flat approximation to the smooth pressure profile seems crude, this is sufficient to describe the effect of the pressure on the global geometry of the equilibrium.
 To illustrate this, we compare the axisymmetric high-pressure equilibrium computed with SPEC (using the stepped pressure profile) to the corresponding VMEC equilibrium (computed with the smooth pressure profile).
 For the VMEC calculation, the smooth rotational-transform profile shown in \Fig{Transform} was used.
 Shown in \Fig{VMECComparison} are the cross-sections of the interfaces as computed by SPEC and the corresponding irrational flux surfaces computed by VMEC.
 That the geometry of the interfaces agrees so well is partly due to the fact that the Shafranov shift is rather insensitive to the pressure profile itself, but depends on the integral of the pressure, i.e. the plasma beta.
 Also, the location of the magnetic axis and the geometry of the innermost interface computed by SPEC agree well with that computed by VMEC.

 In \Fig{VMECComparison} we also show the radial sub-grid resolution used in each annulus.
 In total, there are $78$ global radial degrees of freedom used in the piecewise-quintic representation of each Fourier harmonic of the magnetic vector potential.

 Between the interfaces in the SPEC equilibrium, the rotational transform is not given \emph{a priori}: the rotational transform within each annulus is to be determined as part of the equilibrium calculation.
 An approximation to the global rotational transform may be constructed \emph{a posteori} by field-line following, and this is shown in \Fig{Transform}.
  
 To this high-pressure axisymmetric equilibrium we add a boundary perturbation that resonates with low-order rational surfaces inside the plasma, which induces a large fraction of the equilibrium magnetic field to become chaotic.
 \insertfigure{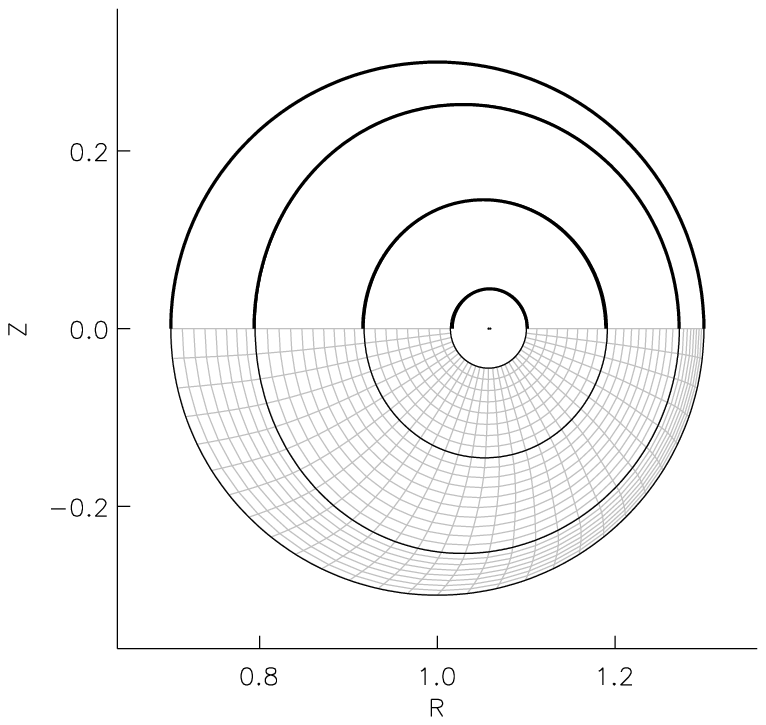}{The ideal interfaces of the high-pressure equilibrium computed by SPEC (thin lines, both upper and lower half) and the corresponding irrational surfaces of the VMEC equilibrium (thick lines, upper half only). In the lower half, the radial sub-grid and the computational angle coordinate is also shown (grey lines).}{VMECComparison}
 While the {\em size} of the magnetic islands created by 3D shaping is determined by the magnitude of the resonant component of the deformation, and the shear, their {\em location} is determined by the rotational-transform profile.
 The rotational transform is specified discretely and is only constrained at the interfaces themselves; however, with the interface rotational transforms given in \Tab{interfacetransforms}, the location of the $\iotabar=1/2$ resonance is guaranteed to lie in the third annulus (that lies between the second and third interfaces).
 Other $n=1$ resonances, such as the $\iotabar=1/3$, $\iotabar=1/4$ and $\iotabar=1/5$ resonances, are guaranteed to lie in the fourth (outermost) annulus.
 The $m=0$, $n=1$ resonance is not present.

 In all of the calculations referred to below, force balance at each interface is satisfied to within \mbox{$|\boldF|<10^{-12}$}, or less.
 
 \subsection{Helical magnetic axis}

 For the zero-beta equilibrium, with $p_0=0$, the fixed outer boundary is described by 
 \be \begin{array}{ccr} R & = & 1.0 + \delta \cos(\zeta) + 0.3 \cos(\theta), \\ 
                        Z & = &       \delta \sin(\zeta) + 0.3 \sin(\theta),
     \end{array} \nonumber
 \ee
 where the helical deformation is \mbox{$\delta = 0.035$}.
 The Fourier representation of both the ideal interfaces and the vector potential in each annulus includes the modes \mbox{$0 \le m \le M$} and \mbox{$-N \le n \le N$}, for $(M,N)=(6,3)$.
 \Poincare plots of the equilibrium are shown on three cross sections in \Fig{Poincarehelical}.
 \insertdblfigure{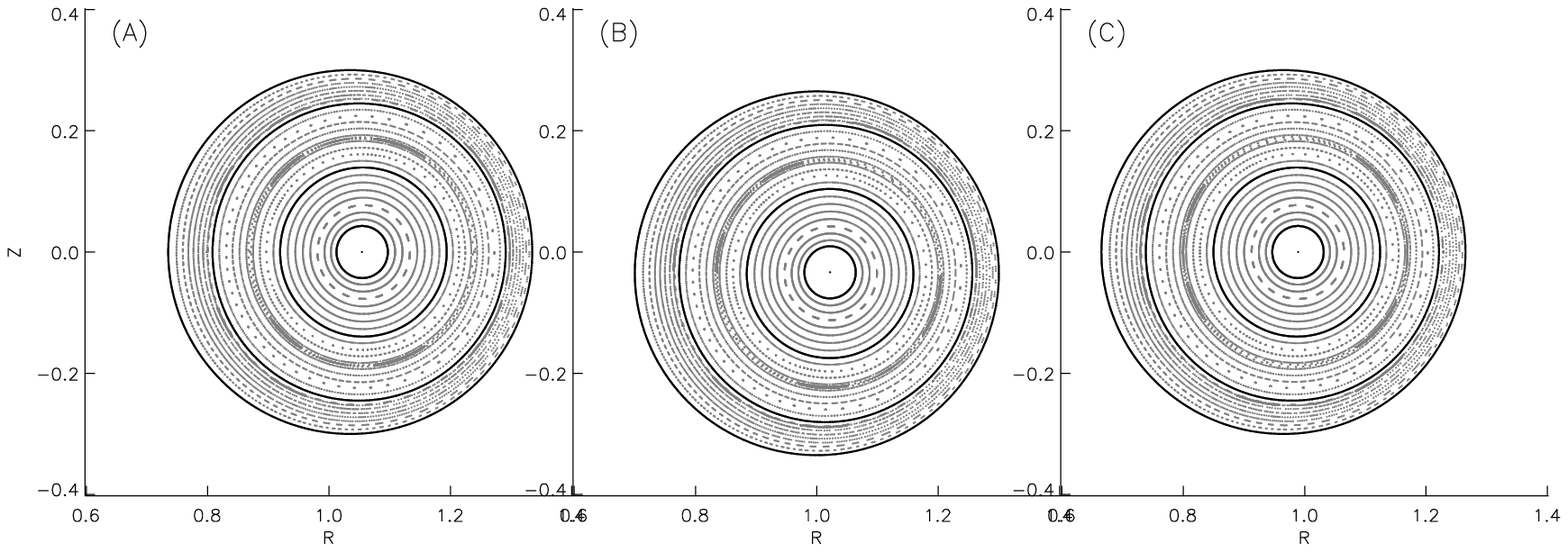}{\Poincare plots (gray dots) of the helical axis equilibrium on (A) $\zeta=0$; (B) $\zeta=\pi/2$; and (C) $\zeta=\pi$. The ideal interfaces are shown with black lines.}{Poincarehelical}

 Nonaxisymmetric configurations are not guaranteed to be integrable, but neither are they guaranteed to be globally chaotic.
 If one were to look closely, small islands may be observed at all the rational surfaces inside the plasma; but in this case, because the nonaxisymmetric deformation of the boundary does not directly resonate with any rational surfaces, no large island chains will form.

 To illustrate an equilibrium that does have a significant volume of chaotic fields, in the next example we include large boundary deformations that resonate with low order rational surfaces.
 
 \subsection{Strongly chaotic equilibrium}

 To drive islands in the high-pressure configuration we include a deformation of the minor radius that resonates directly with the \mbox{$\iotabar=1/2$} and \mbox{$\iotabar=1/3$} rational surfaces.
 The outer boundary is given by 
 \be \begin{array}{ccr} R & = & 1.0 +  \left[ 0.3 + \delta \cos(2\theta-\zeta) + \delta \cos(3\theta-\zeta) \right] \cos(\theta), \\ 
                        Z & = &        \left[ 0.3 + \delta \cos(2\theta-\zeta) + \delta \cos(3\theta-\zeta) \right] \sin(\theta),
     \end{array} \nonumber
 \ee
 where the magnitude of the deformation is given \mbox{$\delta=0.003$}.
 The Fourier representation includes the modes \mbox{$0 \le m \le M$} and \mbox{$-N \le n \le N$}, for $(M,N)=(9,4)$.
 \insertdblfigure{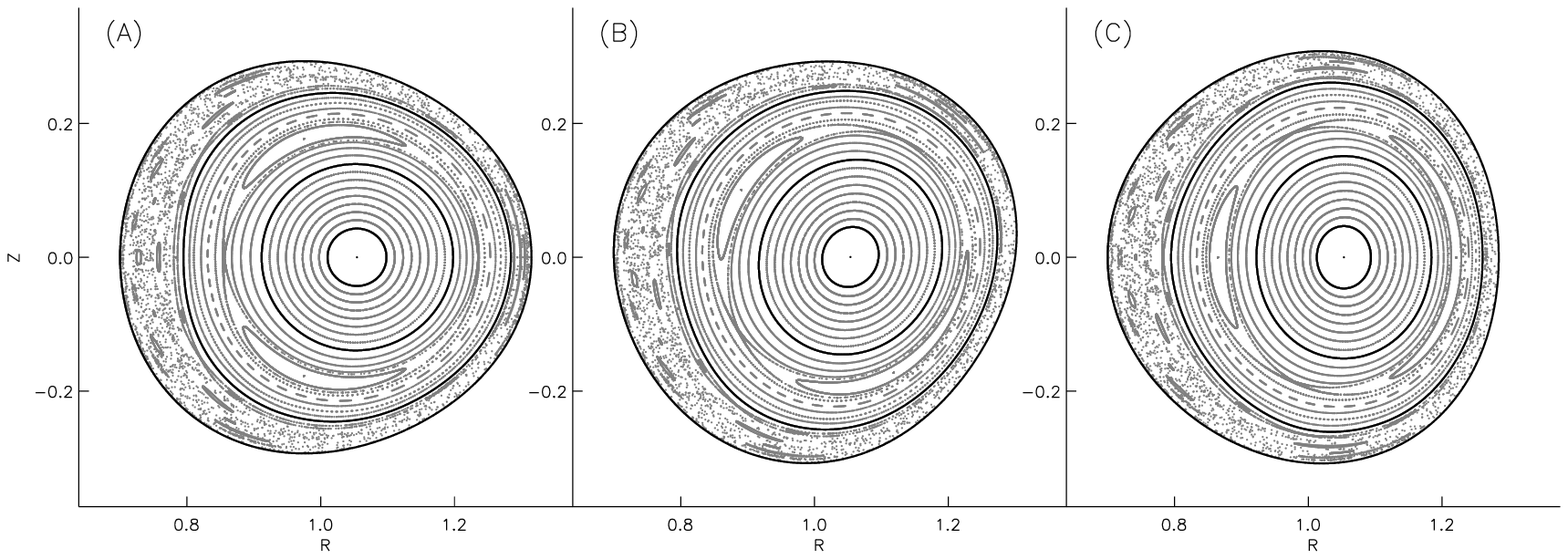}{\Poincare plots (gray dots) of the resonantly deformed equilibrium on the (A) $\zeta=0$, (B) $\zeta=\pi/2$, and (C) $\zeta=\pi$ toroidal cross-sections. The ideal interfaces are shown with black lines.}{PoincareB}
 The field in the outermost annulus is now strongly chaotic.
 This is because this annulus contains several low-order resonances, e.g. \mbox{$\iotabar=1/3$}, which is directly driven by the applied boundary deformation, and also the \mbox{$\iotabar=1/4$}, \mbox{$\iotabar=1/5$}, and \mbox{$\iotabar=1/6$}, and islands will form at these locations islands because of toroidal and poloidal coupling.
 These islands are quite close together and the magnitude of the deformation is sufficient to ensure that these islands overlap.

 To confirm that this solution is converged with respect to Fourier resolution, we recompute this equilibrium using the Fourier resolution $(M,N)=(6,1)$, $(7,2)$, and $(8,3)$, and compare the geometry of the interior interfaces.
 Simply comparing the Fourier harmonics of the interfaces at different resolutions may give misleading results  because, as the Fourier resolution is increased, the spectral condensation algorithm has more opportunity to exploit the tangential freedom, and thus to give a slightly different angle parametrization of the same geometrical interface.
 To eliminate any uncertainty arising from this, we introduce the following angle-independent measure of the geometrical difference between a given pair of interfaces,
 \be d_{M,N}(\theta,\alpha) \equiv \sqrt{ \left[ x(\theta)-x_{M,N}(\alpha) \right]^2 + \left[ y(\theta)-y_{M,N}(\alpha) \right]^2 }, \nonumber
 \ee
 where $x(\theta)\equiv R(\theta,\zeta_0)$ and $y(\theta)\equiv Z(\theta,\zeta_0)$ is the interface cross-section curve of a reference solution (specified below), and $x_{M,N}(\alpha)\equiv R_{M,N}(\alpha,\zeta_0)$ and $y_{M,N}(\alpha)\equiv Z_{M,N}(\alpha,\zeta_0)$ is the cross-section curve of the solution computed with Fourier resolution $(M,N)$, on the plane $\zeta=\zeta_0$.
 We then compute the angle-independent measure of the error according to 
 \be \Delta_{M,N} \equiv \int_{0}^{2\pi} D_{M,N}(\theta) \sqrt{x'(\theta)^2+y'(\theta)^2} d\theta,
 \ee
 where \mbox{$D_{M,N}(\theta) \equiv \min_{\alpha} d_{M,N}(\theta,\alpha)$}.
 Taking the reference configuration to be the solution computed with the highest Fourier resolution, i.e. with $(M,N)=(9,4)$, so that $\Delta_{9,4}=0$ by definition, the convergence error for the interior interfaces is shown in \Fig{Convergence}.

 \insertfigure{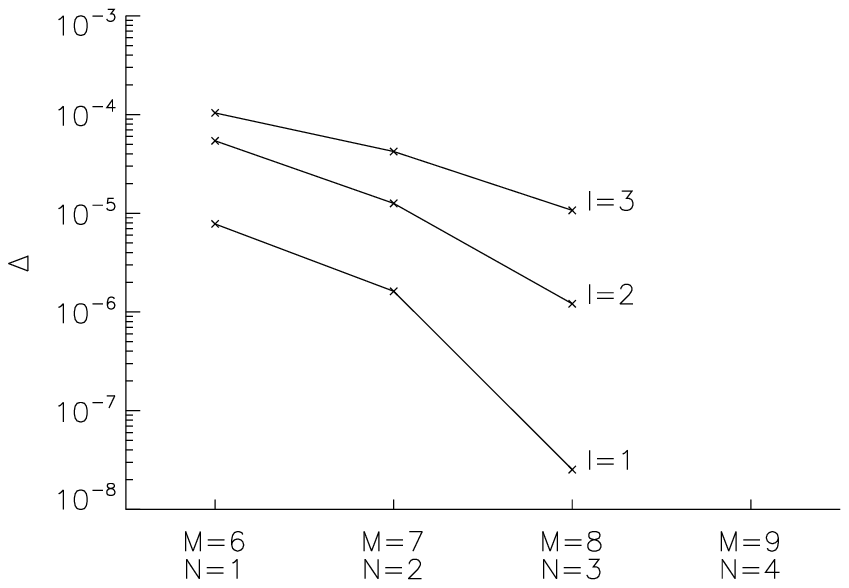}{Angle-independent measure of the error in the interior interface cross-section geometry, with reference configuration $(M,N)=(9,4)$ and $\zeta_0=0$.}{Convergence}

 \section{Comments}

 In this article we have constructed stepped-pressure equilibria with only 4 nested annular regions, as this seems to be sufficient in order to capture the global deformation induced by non-trivial pressure, as confirmed by \Fig{VMECComparison}.
 It is possible to compute equilibria with arbitrarily many interfaces and annular regions; however, as the number of ideal interfaces increases, and the separation between the ideal interfaces decreases, the present modified-Newton algorithm becomes more fragile because of the emergence of small eigenvalues in the matrix, \mbox{$\nabla \boldF$}.
 In future work we hope to explore alternative numerical algorithms for constructing solutions.
 Given that stepped-pressure equilibria are defined as extrema of a constrained energy functional, it should be possible to implement a rapid, preconditioned descent-style algorithm \cite{Hirshman_Betancourt_91}.
 Also in future work, we intend to compute the Beltrami field in the innermost volume (which contains the toroidal coordinate singularity), so that the geometry of the innermost interface can be determined directly from force balance, rather than appealing to regularity conditions at the origin and employing extrapolation methods.

 Specifying the profiles discretely may seem arbitrary, but it is a practical means of maintaining some \emph{a priori} control over the pressure and rotational transform while making minimal assumptions regarding the topology of the field.
 In fact, the only topological assumption made is that, where there are pressure gradients there must be irrational flux surfaces.
 Thus the prescription of the profiles may be made arbitrarily fine-grained as long as magnetic surfaces exist.
 Where they do not exist, force-free fields are the only choice consistent with no mass flow, Beltrami fields being the most practical. 

 To conclude, we make some comments regarding the existence of MHD equilibrium solutions, and the ``pathological'' nature of the pressure profile.
 Earlier we commented how solutions to $\nabla p = {\bf j}\times{\bf B}$ with non-integrable magnetic fields with a fractal phase space must have pressure profiles with infinitely discontinuous gradient.
 In the MRXMHD model, the pressure profile is piecewise flat, and possibly discontinuous at the ideal interfaces.
 Such a profile may also be described as pathological; however, the MRXMHD model is based on an integral principle, and a discontinuous pressure profile remains an integrable function, provide the number of discontinuities is finite, as it certainly is in the above calculations.
 The MRXMHD equilibrium is well defined mathematically, and at no point in the numerical construction of the stepped-pressure equilibrium is the pressure gradient required.

 In the analysis of the force-balance condition, \mbox{$[[p+B^2/2]]=0$}, arising in the Euler-Lagrange equation, \Eqn{firstvariation}, it was shown \cite{MHDvN_10} that generally pressure can only be supported if the interfaces have irrational rotational transform.
 This in turn places constraints on the pressure and rotational-transform profiles that are used to define the equilibrium: if pressure is placed on the rational interfaces, then no equilibrium solution will exist.

 An analogous condition holds for ideal, scalar pressure equilibria with nested flux surfaces, i.e. integrable magnetic fields.
 States that minimize the plasma energy, $U$, allowing for ideal variations, must satisfy the Euler--Lagrange equation, $\nabla p ={\bf j}\times{\bf B}$, which is the analogous statement of force-balance for ideal MHD equilibria with nested flux surfaces, i.e. ideal equilibria that are globally topologically constrained, rather than discretely topologically constrained as in MRXMHD.
 An analysis of this equation shows that there is a singularity in the resonant harmonic of the parallel current at each rational surface \cite{BHHNS_95}, which are dense in any system with shear.
 In addition to the ideal $\delta$-function surface currents required to shield resonant fields, that would otherwise result in the formation of magnetic islands, there generally exist pressure-driven $1/x$ style singularities, where $x\equiv (\iotabar-n/m)$, which  are required to satisfy quasi-neutrality.
 Writing the current as ${\bf j}=\sigma {\bf B}+{\bf j}_\perp$, and insisting that $\nabla \cdot {\bf j}=0$, the parallel current, $\sigma$, must satisfy ${\bf B}\cdot \nabla \sigma = - \nabla \cdot {\bf j}_\perp$, where the perpendicular current is given by force balance.
 The magnetic differential equation is singular, and solvability conditions on the perpendicular current must be satisfied if a single valued $\sigma$ is to exist\cite{Newcomb_59}: an arbitrary ${\bf j}_\perp = {\bf B}\times \nabla p / B^2$ is not consistent with quasi-neutrality!
 The singularity in the ${\bf B}\cdot \nabla$ operator is exposed by employing straight-field-line coordinates, which can be constructed globally if, and only if, the magnetic field is integrable, so that $\sqrt g {\bf B}\cdot \nabla=\partial_\zeta + \iotabar \partial_\theta$.
 The solvability condition that must be satisfied for quasineutrality is that, in arbitrary geometry, the pressure gradient must go to zero at each rational surface at least as fast as $(\iotabar-n/m)$, and at least as fast as $(\iotabar-n/m)^2$ if the pressure is to remain monotonic.
 Given that the rational surfaces are dense (i.e. arbitrarly close to any point in space) in plasma equilibria with shear, this results in a pressure profile that may also be described as pathological.

 The MRXMHD approach seeks integrable solutions, rather than differentiable solutions.
 The philosophy of seeking weak solutions was endorsed by Garabedian, who claimed that ``differentiable solutions of the equilibrium equations do not exist in general when the geometry is three-dimensional, so that weak solutions are required to model the physics adequately'' \cite{Garabedian_98}.
 

 \end{document}